\documentclass{aa}
\usepackage{txfonts}
\usepackage{graphicx}
\usepackage{natbib}
\usepackage{mathptm}
\usepackage{psfig}
\begin{document}

\title{Orbital Parameters of Merging Dark Matter Halos}
\author{S. Khochfar\inst{1} \and A. Burkert\inst{2}}

\institute{Department of Physics, 
Denys Wilkinson Building, Keble Road, Oxford OX1 3RH, United Kingdom 
\and University Observatory Munich, Schreinerstr. 1, 81679 Munich, Germany 
}
\date{Received / Accepted }

\abstract{
In order to specify cosmologically motivated initial conditions for major galaxy mergers
(mass ratios $\leq$ 4:1)
that are supposed to explain the formation of elliptical galaxies we study the
orbital parameters of major mergers of cold dark matter halos
using a high-resolution cosmological simulation. 
Almost half of all encounters are nearly parabolic with eccentricities $e
\approx 1$ and no correlations between
the halo spin planes or the orbital planes. The pericentric argument
$\omega$ shows no correlation with the other orbital parameters and is
distributed randomly. In addition we
find that 50 \% of typical pericenter distances are larger than half the halo's virial radii which is
much larger than typically assumed in numerical simulations of galaxy
mergers. In contrast to the usual assumption made in semi-analytic
models of galaxy formation the circularities of major mergers are
found to be not randomly distributed but to peak around a value of
$\epsilon \approx 0.5$. Additionally all results are independent of
the minimum progenitor mass and major merger definitions (i.e. mass
ratios  $\leq$ 4:1; 3:1; 2:1).
}

\titlerunning{Merging Dark Matter Halos}
\authorrunning{Khochfar \& Burkert}
\maketitle

\section{Introduction}\label{chorb}

 \citet{TT72} (TT72) proposed that elliptical galaxies could form from
mergers of disk galaxies. Thereafter this "merger hypothesis" has been investigated
in great details by many authors (for recent reviews see Burkert \& Naab 2003a,b) using numerical
simulations. The first fully self-consistent models of two equal-mass stellar disks embedded
in dark matter halos were performed by \citet{neg83}, \citet{barn88} and
\citet{her92}. Typically, the spiral galaxies were constructed in dynamical
equilibrium \citep{her93a} and approached each other on parabolic orbits with pericenter
distances of twice the scale length of the larger disk galaxy. Already the first
simulations demonstrated that mergers lead to slowly rotating, pressure supported
and anisotropic systems that generally follow a de Vaucouleurs surface density profile
and resemble elliptical galaxies. \citet{her93b} noted that
a massive central bulge component was required in the progenitors in order to
explain the high stellar densities observed in the cores of ellipticals.
More recent investigations showed that the initial mass ratio
of the merging spirals determines the isophotal shapes and kinematical properties
of the merger remnants \citep{bendo00} and 
it was suggested e.g. by \citet{naa99} that the initial mass ratio 
might explain the observed dichotomy between disky, fast rotating and
boxy, slowly rotating  ellipticals \citep{bend88,bende88}.

Recently, \citet{naa03} performed the to date largest parameter survey of collisionless
binary disk mergers with varying mass ratios, different initial disk orientations
and orbital parameters. They found that the fine structure of the merger remnant in general depends
strongly on the adopted initial conditions. Certain initial conditions in fact lead to disky, anisotropic
remnants which are not observed. Clearly, a deeper understanding
of the origin of the early type family of galaxies not only requires a more sophisticated
treatment of the various physical processes but also a better
understanding of the 
initial orbital configurations of the interacting galaxies as predicted by currently
favoured cosmological models.

Up to now due to the lack of knowledge of
the appropriate initial conditions, simulations were  forced to cover a wide
parameter space by setting up mergers with different orbital
configurations. This approach also has the drawback that it is not
clear how relevant a given parameter combination is which leads to a certain fine structure
of the merger remnants.
In this paper we therefore present a detailed analysis of a high-resolution, large-scale
cosmological simulation  that was
 carried out by the Virgo Supercomputing Consortium using computers based at Computing Centre of the Max-Planck Society in Garching and at the Edinburgh Parallel Computing Centre,\footnote{ www.mpa-garching.mpg.de/NumCos} 
  and derive
the typical orbital parameters predicted by cosmology for merging dark
matter halos which should also be representative for their visible
galactic components. 

\subsection{The cosmological simulation}

The simulation we analyse was  performed within the GIF project \citep{kau99}  and we will here 
focus on the $\Lambda$CDM simulation.  This simulation was carried out in a box of size
141.3 Mpc $h^{-1}$ with $256^3$ particles each having a mass of $1.4 \times
10^{10} M_{\odot} h^{-1} $ and cosmological parameters $\Omega_{\Lambda}=0.7$,
$\Omega_{0}=0.3$, $\sigma_8=0.9$, $h=0.7$ and a force softening of 
\linebreak
{$20$ Kpc/$h$}. The softening leads to an error of a few percent in the calculation 
 of the orbital parameters.
The positions and velocities of the particles
have been  saved at 44 different redshifts.
 Additionally, for each
output redshift, a list of halo properties have been provided by the GIF project.
When two halos approach each other their orbit is going to change due to the
transfer of orbital angular momentum to the halo's internal angular momentum
which in
the following is called spin and should not be confused with the cosmological spin parameter
defined e.g. in \citet{peeb93}.
The determination of the initial orbital parameters therefore is a question of the
'right timing'. We try to identify these parameters at a time when the
interaction between the halos is still weak and one can assume a Keplerian two-body
situation, using the positions of the most bound
particles of each halo. At each redshift we go through the list of
halos identified by using the friends-of-friends (FOF) algorithm and identify the
positions of the most bound particles. If at one redshift a halo has
disappeared through merging with another halo, we look up the position of its
most 
bound particle at the previous redshift and check whether the distance to the
most bound particle of the other halo, with which it is going to merge,
is larger than the 
sum of both virial radii. If so, we  derive the orbital informations using
the data from this redshift. To make sure that the merger is not
just a flyby of an unbound halo we check at a redshift later
than the merging redshift whether the separation of the
most bound particles has increased again.
Since outputs are generated roughly every $0.5$ Gyr  we expect not to miss
even very eccentric orbits.

\section{The reduced two-body problem}
The orbital parameters are defined as specified in the 
pioneering work of TT72.
We simplify the problem by reducing the two-halo system to a two-body system with
each halo being represented by its most bound particle which sits in the
potential minimum and is expected to be  the most 'stable' particle, allowing us to
follow the orbital evolution of the progenitor halos during the early stages of the
merger even when a clear definition of the centre of mass of the extended and already
partly overlapping progenitor halos is not possible anymore. 
 We checked that the position of the most bound particle is
not very different from the centre of mass of each halo when the halos are
well separated and that it agrees with the region of highest dark matter
density. We additionally tested if the results presented in this paper
show any significant dependence on using the centre of mass of each
halo, calculated by averaging the particle positions inside different
fractions of the virial radius, instead of the position of the most
bound particle and found no significant difference. 
 The position of the most bound particle in 
each of the halos will be denoted by $r$ with the index $h$, like halo, 
for the more massive of the two
merging halos and index $s$, like satellite, 
for the less massive halo. Throughout this paper we will use these 
indexes to refer quantities to the more and less massive of the merging halos.

 In the following we will assume that each of
the two point masses  moves with the average velocity
of the particles inside the virial radius of the corresponding halo.

 Knowing the average velocity and mass inside the virial radii of the merging halos it is possible
to solve the gravitational two body  problem analytically by making the transition to the equivalent
 one-body problem of the reduced mass system \citep[e.g.][]{gol02}. The equation of motion reads
 \begin{eqnarray}\label{emo}
   \mu {\bf \ddot r }=-G \frac{M_h M_s}{r^2} \frac{{\bf r}}{r}.
 \end{eqnarray}
with $M_h$ and $M_s$ being the mass inside the virial radius of the more and less massive of the 
merging halos, respectively. The reduced mass is defined by
\begin{eqnarray}
  \mu \equiv \frac{M_h M_s}{M_h+M_s}.
\end{eqnarray}
Equation \ref{emo} describes the motion of a fictitious particle of mass $\mu$ moving around a point
at a distance $r=|r_h-r_s|$. To specify the orbit we calculate the total energy and orbital 
angular momentum according to
\begin{eqnarray}
  E=\frac{1}{2} \mu \dot{r}^2 -\frac{G M_h M_s}{r}
\end{eqnarray}
and 
\begin{eqnarray}
  L=\mu \bf{r} \times {\bf \dot{r}},
\end{eqnarray}
respectively.
In this study we also investigate the pericenter distance $r_{peri}$ and the eccentricity $e$
of orbits, which we can readily calculate from the energy and orbital angular momentum by
\begin{eqnarray}
  e=\sqrt{1+\frac{2E L^2}{\mu (G M_h M_s)^2}}
\end{eqnarray}
and
\begin{eqnarray}
  r_{peri}=\frac{L^2}{(1+e)\mu G M_h M_s}.
\end{eqnarray}  
In the following we will identify orbits according to their total energy and eccentricity into
hyperbolic ($E>0$, $e>1$), parabolic ($E=0$, $e=1$) 
and elliptic ($E < 0$, $e < 1$) orbits.

\subsection{Limits on the two-body approximation}

\begin{figure}
\begin{center}
\includegraphics[width=8cm,angle=0]{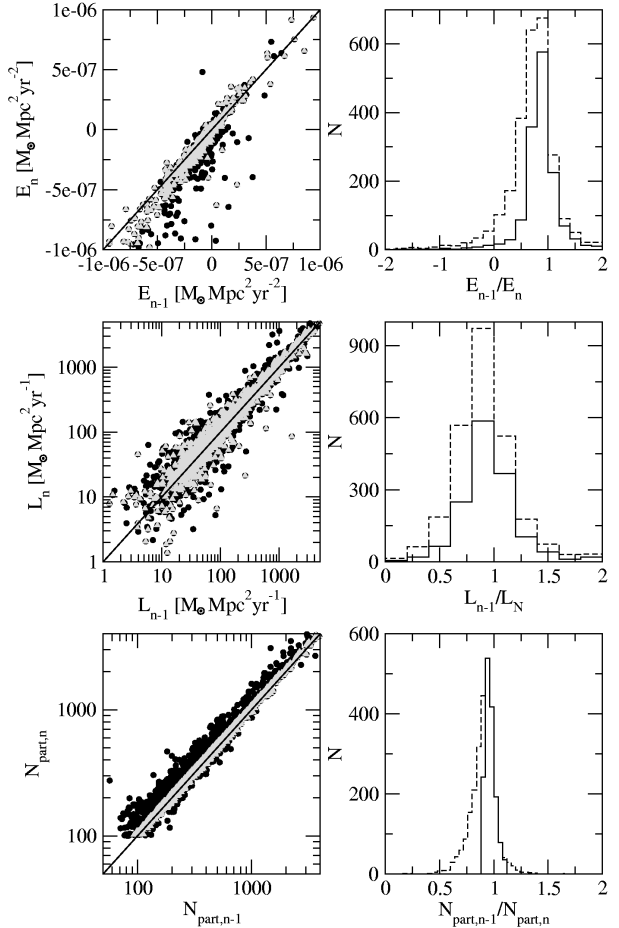} 
\medskip
\caption{Upper left panel: Total energy of the two halo system. 
Upper right panel: Distribution of ratios $E_{n-1}/E_{n}$.
 Middle panel: Orbital angular momentum of the two halo system. 
Middle right panel: Distribution of ratios $L_{n-1}/L_{n}$.
Lower panel: Mass of the progenitors in units of particle mass in 
the simulation. Lower right panel: Distribution of ratios 
$N_{part,n-1}/N_{part,n}$.
The indexes $n$ and $n-1$ refer to one and two simulation time steps 
before the merger. The circles show all  identified mergers, and the 
triangles only those when applying the mass  criterion. Dashed histograms 
show the distribution for all mergers and solid histograms only for those 
satisfying the mass criterion.
  \label{ch1}}  
\end{center}
\end{figure}

As summarised in the previous section we assume the two merging halos 
to be fairly 
isolated and without experiencing any other severe interactions 
before they merge. 
This however, is not guaranteed  
and needs to be checked. We calculated the total energy and 
the orbital angular 
momentum of the orbits of 
the two halos at two simulation outputs before they merge. To do so 
we followed the most massive 
progenitors backward in time and analysed their orbital parameters, 
including all halos with more than 50 particles and with a mass ratio of  
4:1 or less. 
 The filled circles in top left panel of Fig. \ref{ch1} 
show the total energy of 
the merging halos $E_n$ and $E_{n-1}$, one and 
two time steps before the merger, respectively. As can be seen the energies 
seem to be not too different between the two time steps, with a tendency for 
bound orbits at time step $n-1$ to become more bound at $n$. The dashed 
histogram in top right panel shows the distribution of 
ratios $E_{n-1}/E_{n}$ for all identified mergers. The distribution is 
skewed toward values smaller than one which confirms the tendency of the 
orbits to become more bound between two time steps. The filled circles in the 
left middle panel of the same figure shows the correlation 
between the orbital angular momentum at two preceding time steps before the 
merger. The dashed histogram in the right middle panel shows the distribution
of the ratio of these orbital angular momenta. In general the orbital angular 
momentum between two outputs does not change. However we 
find that there is a considerable scatter which is symmetrically distributed 
around $L_{n}=L_{n-1}$.

 The aim of this work is to present initial conditions 
for high resolution simulations of isolated merging galaxies. 
 This means that we ideally would like the energy and angular 
momentum to be conserved or not to change significantly while 
the two merging halos are well separated on their approach.
As can be seen in the lower left panel of Fig. \ref{ch1} many of the 
progenitors are not isolated in the sense that they do not accrete further 
mass before they merge. To choose fairly isolated halos we therefore 
decide to only consider those mergers in which 
$ 1.11 (M_{h,n}+M_{s,n}) \geq M_{h,n-1}+M_{s,n-1} \geq 0.9 (M_{h,n}+M_{s,n})$. 
The result of this selection criterion is shown by the grey triangles in the 
left panels of Fig. \ref{ch1}. In all cases the correlations between the 
quantities at the different redshifts become tighter. In addition we find that
the distribution in the ratio of energies is now centred around a value of 
one (solid histogram right top panel). The shape of the distribution in 
the ratios of the angular momenta does not change much though it becomes a  
bit narrower. In the following we will use above mass criterion to select 
mergers for the sample used here. This criteria roughly rejects $ 50\%$ of the 
mergers detected in the simulation.

\section{Orbital parameters $r_{peri}$ and $e$ for major mergers }

We start by analysing the dependencies and correlations of the pericenter
distance $r_{peri}$ and the eccentricity $e$  of merging halos extracted
from the simulation data  at one output prior to the merger.  
We restrict ourselves to mergers with mass ratios of 
4:1 or less as numerical models have shown that larger mass ratios do not resemble
elliptical galaxies anymore as the more massive disk is not destroyed
\citep{barn92,naa03}.

\begin{figure}
\begin{center}
\includegraphics[width=8cm,angle=0]{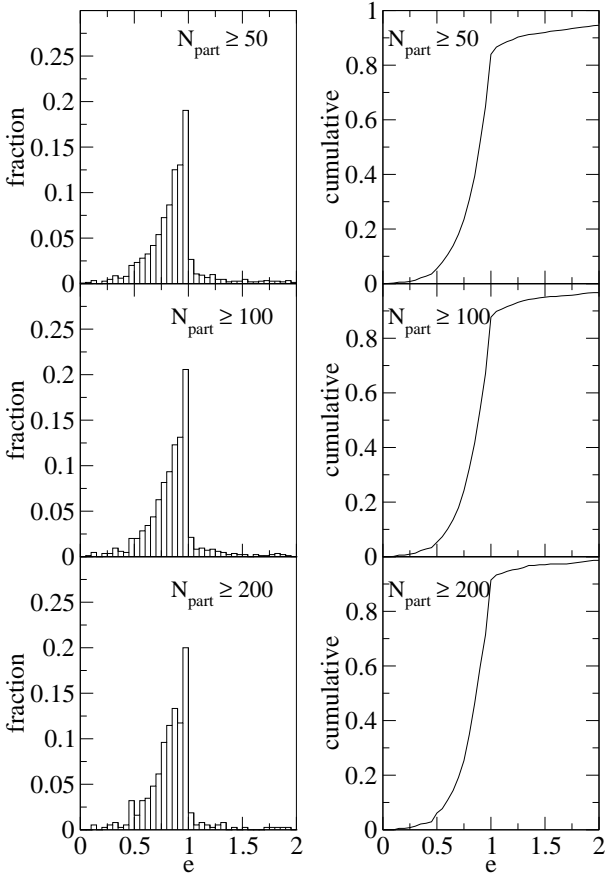} 
\medskip
\caption{ Left panel: Eccentricity distribution of merging halos with mass larger than
  $N_{part} \cdot 2 \times 10^{10} M_{\odot}$.
  $N_{part}$ is the number of dark matter particles.
 Right panel: Corresponding
  cumulative fraction of eccentricities.     
  \label{ecc_1}}  
\end{center}
\end{figure}

\begin{figure}
\begin{center}
\includegraphics[width=8cm,angle=0]{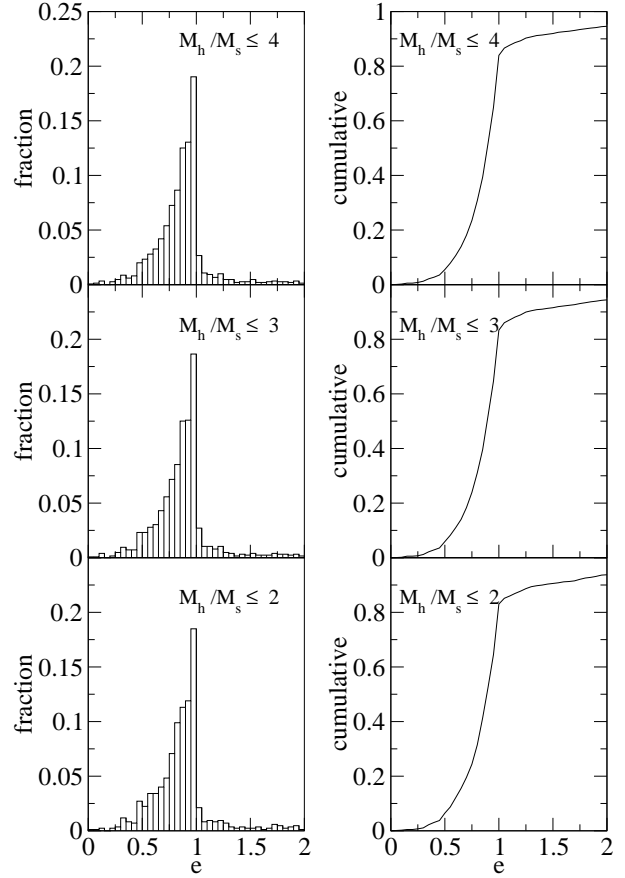} 
\medskip
\caption{ Left panel: Fraction of merging halos with mass larger than
  $ 10^{12} M_{\odot}$ and different mass rations $M_h /M_s$, on
  initial orbits with 
  eccentricity $e$. Right
  panel: Corresponding 
  cumulative fraction of eccentricities.     
  \label{ecc_1b}}  
\end{center}
\end{figure}

\begin{figure}
\begin{center}
\includegraphics[width=8cm,angle=0]{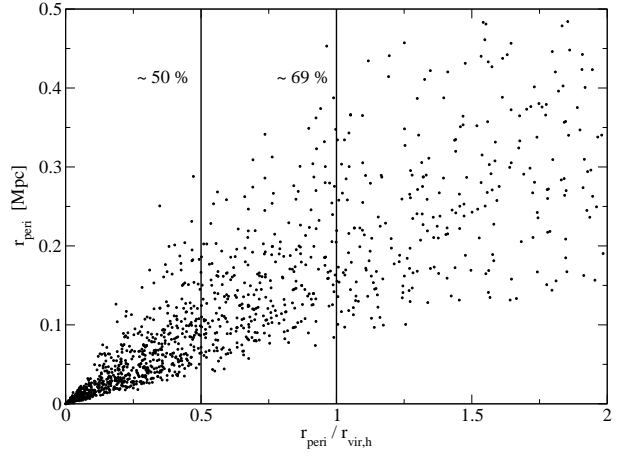}
\medskip
\caption{ Pericenter distance in units of Mpc against pericenter distance in
  units of $r_{vir,h}$. The values at each line indicate the completeness
  limit, meaning the fraction of mergers with $r_{peri}/r_{vir,h} < 0.5$ and
  $1$. Results are shown for progenitor halos of mass larger than 
 \mbox{$ 10^{12} M_{\odot}$} and mergers with mass ratio $M_h /M_s
  \leq 4$.  
  \label{ecc_3}}   
\end{center}
\end{figure}

In figure \ref{ecc_1} the distribution of eccentricities $e$  of merging halos
as function of the minimum mass of the  progenitor halos is shown. 
Roughly half of the orbits are 
parabolic or very close to parabolic. We find $\sim 40 \%$ 
of the orbits in
the range $e=1 \pm 0.1$. This result is independent of the minimum mass cut
applied. N-body simulations of merging galaxies assume parabolic orbits \citep[e.g.][]{barn88},
which is only the case for half of the mergers we find.
In fig. \ref{ecc_1b} we show the dependence of our results on the mass ratio of the
major mergers. We find no dependence with the mass ratio, which indicates a self-similarity of the
formation process. As many characteristic properties of elliptical galaxies as
 e.g. the amount of rotation
are determined by the mass ratio (Naab \& Burkert 2003) the cosmological models 
would predict that massive
and low-mass ellipticals should have similar properties which is not in agreement with observations.
In contrast with the theoretical predictions luminous ellipticals are preferentially anisotropic and
slowly rotating while less luminous ellipticals rotate fast and appear isotropic (Bender 1988). 
\begin{figure}
\begin{center}
\includegraphics[width=8cm,angle=0]{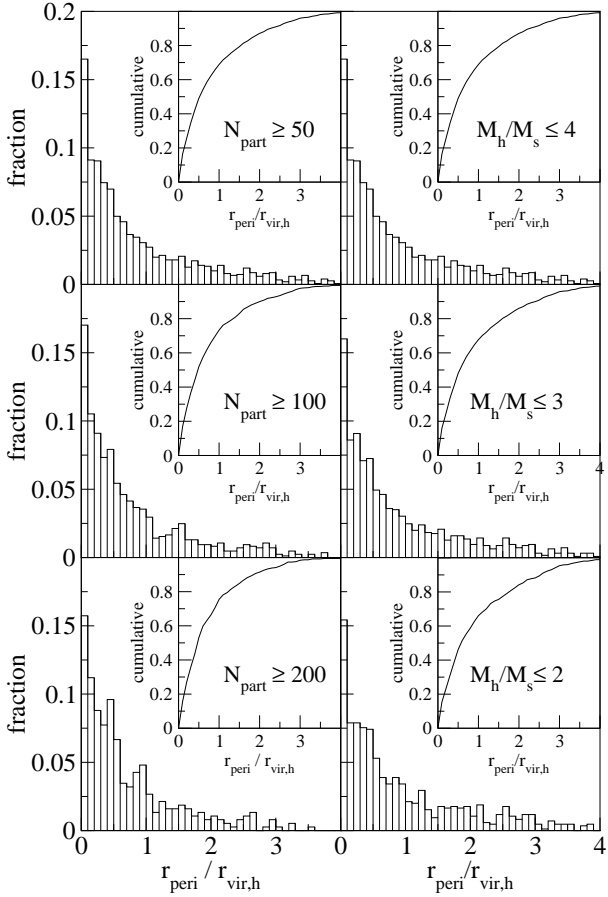} 
\medskip
\caption{  Left panel: Fraction of merging halos with different pericenter 
distances in units of the more massive progenitor's  
virial radius $r_{vir,h}$ and with progenitor masses larger than
  $N_{part} \cdot 2 \times 10^{10} M_{\odot}$. The corresponding
  cumulative fraction of pericenter distances are also shown. All results are 
shown for mergers with a mass ratio of $M_h /M_s \leq 4$.
 Right panel: Fraction of merging halos with mass larger than
  $10^{12} M_{\odot}$ and different mass ratios  $M_h /M_s$, on
  orbits with different pericenter  distances in  units of the 
  more massive progenitor's  virial radius 
  $r_{vir,h}$. Cumulative fractions are shown too.
  \label{ecc_2}}   
\end{center}
\end{figure}
The distribution  of the pericenter distance in units of the virial radius of
the more massive progenitor halo and its dependence on minimum
mass is shown in  the left panels of fig. \ref{ecc_2}. 
 The distribution shows no significant dependence on the minimum mass.
 Small pericenter are more frequent than larger
ones. This is actually what one would expect, because halos
which are on orbits leading to a very close encounter are more likely to merge
than those which pass each other from very far. Roughly  85 \% 
of the mergers
have pericenter distances which are larger than $0.1 r_{vir,h}$.  
 Fig. \ref{ecc_2} show that the
dependence on the mass ratio of the mergers is negligibly small. 
Merger simulations usually adopt initial conditions with small
pericenter distances which are typically of order $0.05 r_{vir,h}$, leading to fast merging
that saves computational time. This assumption is not in agreement with cosmological models
and might affect the inferred structure of merger remnants using numerical simulations.
According to the reduced two-body problem  $r_{peri}\propto
L^2$, which means that the orbital angular momentum in  typical merger simulations
is less than that for merging halos in self-consistent cosmological
simulations. Less angular momentum will therefore be transferred during the merger 
process and the structure of the remnant could be different. 
 In fig. \ref{ecc_3} we correlate the pericenter distance in units of Mpc 
to the pericenter distance
in units of $r_{vir,h}$. We find that on average roughly $50 \%$ 
of all mergers have pericenter 
distances less than $0.5 r_{vir,h}$. About $70 \%$ of all mergers 
lead to a first
pericenter passage which penetrates the virial radius of the larger halo $r_{vir,h}$. 

\begin{figure}
\begin{center}
\includegraphics[width=8cm,angle=0]{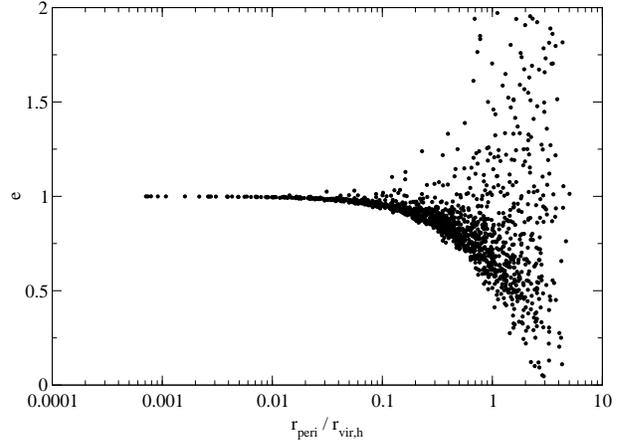}
\medskip
\caption{ Correlation between the eccentricity and pericenter distance of
  merging halos found in the simulation. Results are shown for progenitor
 halos of mass larger than  \mbox{$10^{12} M_{\odot}$} and mergers
 with mass ratio $M_h /M_s \leq 4$.  
  \label{ecc_4}}   
\end{center}
\end{figure}

Since we have found no dependence on the minimum mass and on the mass ratio
we continue our investigations using as a standard  assumption
$M_{min}= 10^{12} M_{\odot}$ which corresponds to the typical halo
size of  massive galaxies and $M_h /M_s \leq 4$.
Not every random orbit is going to lead to a merger and it is
important to see if a preferred orbit configuration exists leading to
mergers. In fig. \ref{ecc_4} we correlate $r_{peri}$ and $e$.
Mergers with $r_{peri} \leq 0.1
r_{vir,h}$ are almost all on parabolic orbits with $e \sim 1$. Orbits with
 $r_{peri} > 0.1 r_{vir,h}$ have a scatter in $e$ which increases with pericenter
 distance toward bound orbits. The same behaviour is found looking at the correlation between
 eccentricities and pericenter distances in units of Mpc. 

\begin{figure}
\begin{center}
\includegraphics[width=8cm,angle=0]{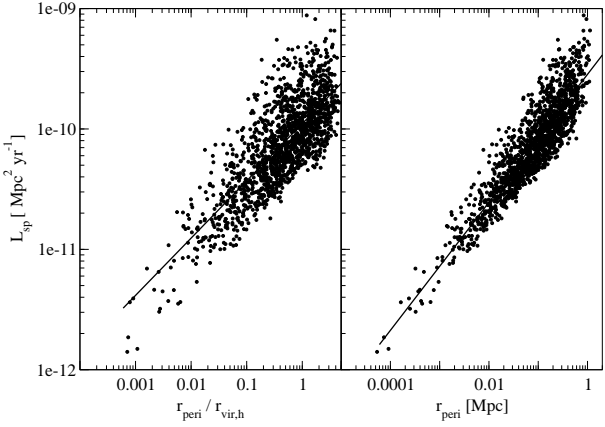}
\medskip
\caption{Correlation between the specific angular momentum $L_{sp}$ of the
  orbit and its 
  pericenter distance in units of $r_{vir,h}$ (left panel) and Mpc (right
  panel). Lines represent power law fits to the data. Results are shown for
  progenitor halos of mass larger than  
 \mbox{$ 10^{12} M_{\odot}$} and mergers with mass ratio $M_h /M_s
  \leq 4$.  
  \label{ecc_5}}   
\end{center}
\end{figure}

 The expectation from the two-body solution is that the specific angular momentum  $
 L_{sp}\equiv L/\mu$  for remnants of the same 
mass $M_h +M_s$ is proportional to $r_{peri}^{1/2}$  with a scatter
 because of  different eccentricities of the orbits. In fig. \ref{ecc_5} we
 show the correlations found between these two quantities. The lines in the
 left and right panel of fig. \ref{ecc_5}  are
 power law fits to the data with 
\begin{eqnarray}
L_{sp}=1.11 \times 10^{-10} \left(\frac{r_{peri}}{r_{vir,h}} \right)^{0.48}
 \frac{\mbox{Mpc}^2}{\mbox{yr}} 
\end{eqnarray}
for the pericenter distance in units of $r_{vir,h}$ and
\begin{eqnarray}
L_{sp}=2.84 \times 10^{-10} \left( \frac{r_{peri}}{\mbox{Mpc}} \right)^{0.53}
 \frac{\mbox{Mpc}^2}{\mbox{yr}} 
\end{eqnarray}
for the pericenter in units of Mpc. The fits show that the data is following
 the trend of $L_{sp} \propto r_{peri}^{1/2}$. The larger scatter in the
 correlation with $r_{peri}$ in units of $r_{vir,h}$ is due 
to  additional 
scatter introduced on the horizontal axis by the variation in the values of 
$r_{vir,h}$ at a given $r_{peri}$.\\ 

\begin{figure}
\begin{center}
\includegraphics[width=8cm,angle=0]{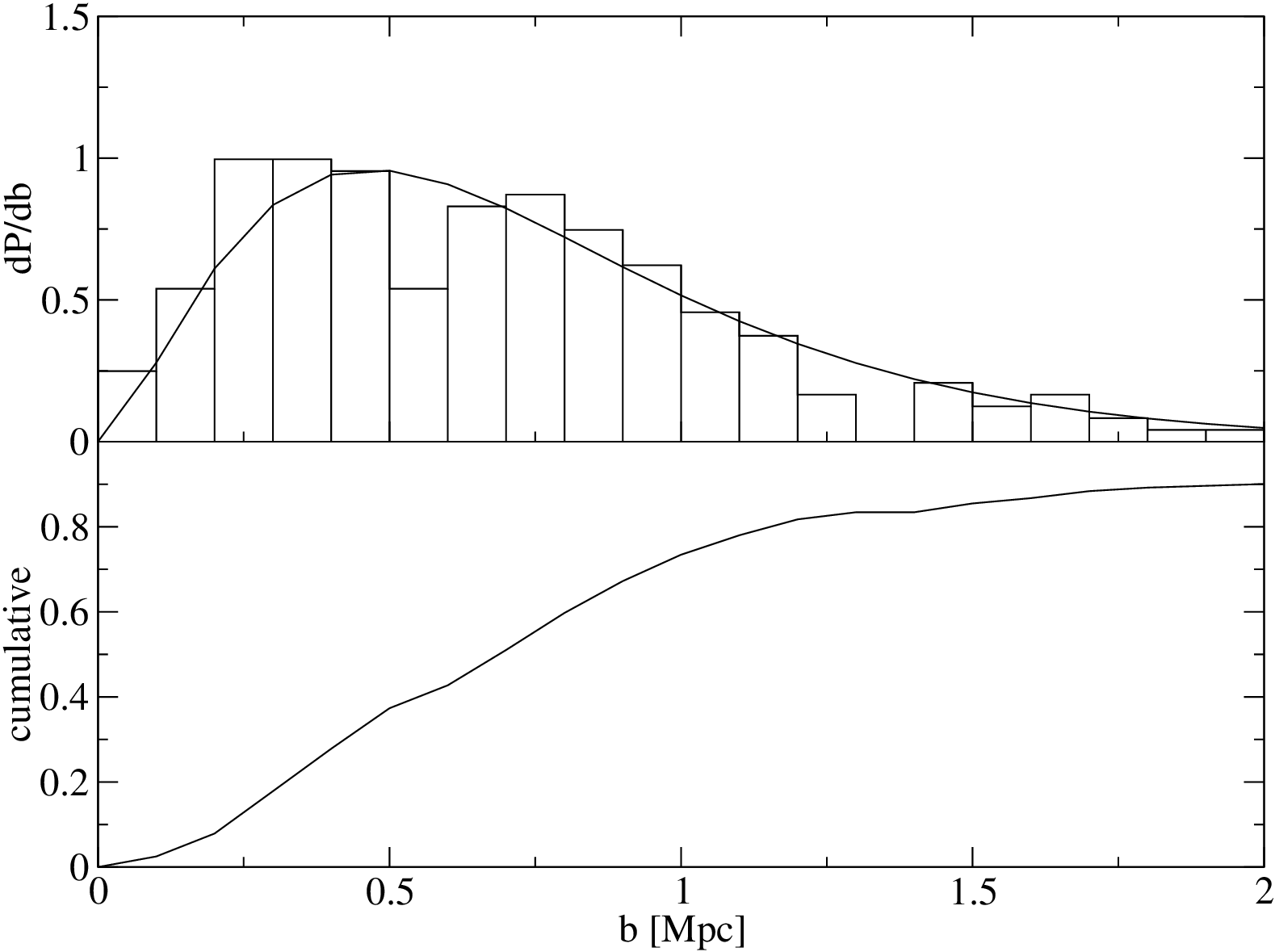}
\medskip
\caption{ Upper graph: Probability density  of hyperbolic orbits with
  different impact 
  parameter $b$. Lower graph: To the upper graph corresponding cumulative
  fraction.  Results are shown for progenitor halos of mass larger than  
 \mbox{$ 10^{12} M_{\odot}$} and mergers with mass ratio $M_h /M_s \leq 4$. 
  \label{ecc_6}}   
\end{center}
\vspace{1cm}
\begin{center}
\includegraphics[width=8cm,angle=0]{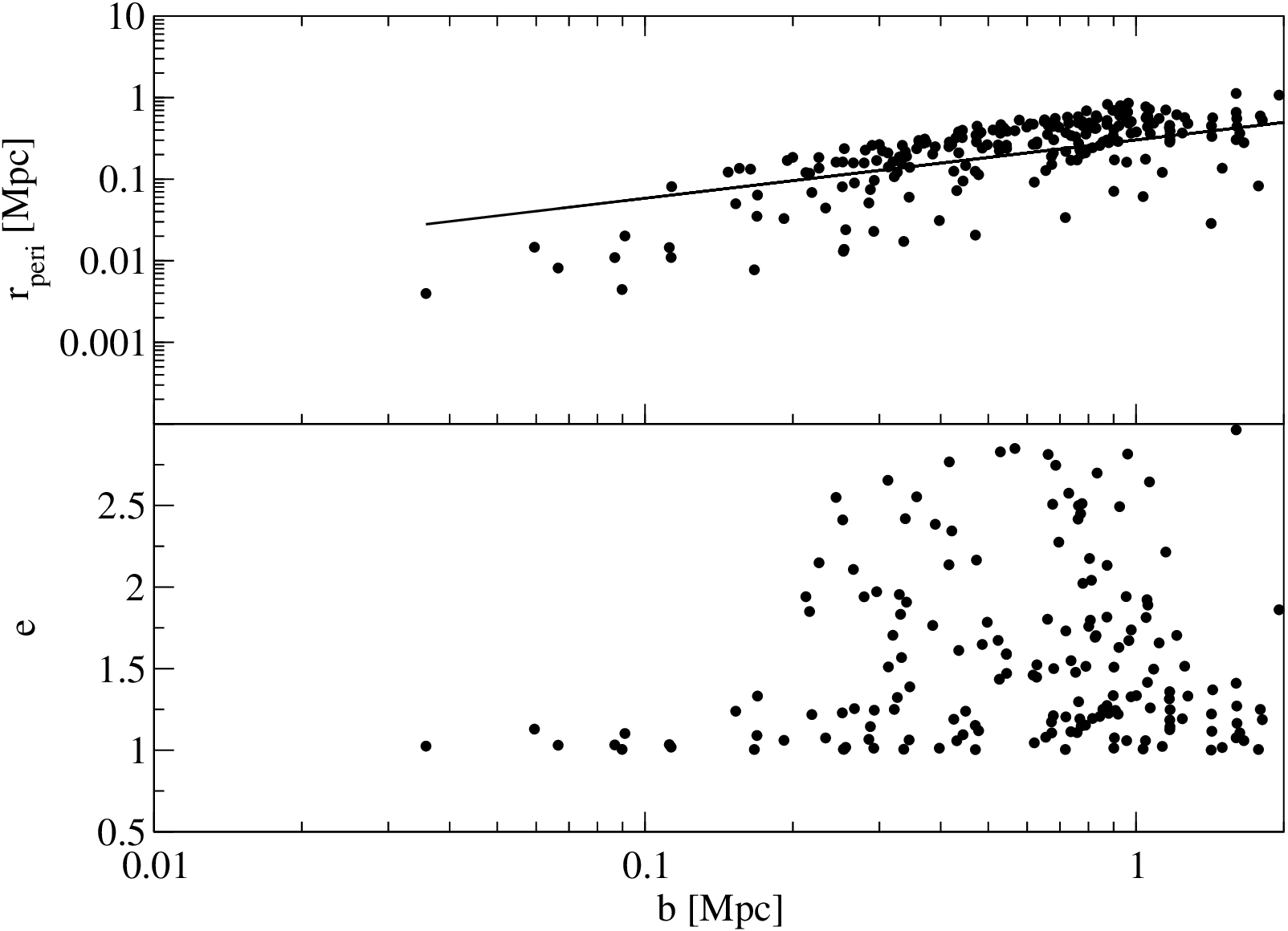}
\medskip
\caption{ Upper graph: Correlation between pericenter distance and impact
  parameter of hyperbolic orbits leading to mergers. Lower graph: Correlation
 between the eccentricity and impact parameter corresponding to the orbits in
 the upper graph.  Results are shown for progenitor halos of mass larger than  
 \mbox{$ 10^{12} M_{\odot}$} and mergers with mass ratio $M_h /M_s \leq 4$. 
  \label{ecc_7}}   
\end{center}
\end{figure}

Hyperbolic encounters are often characterised by the 
impact parameter $b$.
 The impact parameter is defined by the spatial separation perpendicular 
to the initial
 velocity $\bf{V}(t= -\infty)$ which has the value
\begin{eqnarray}
V_{-\infty}=\sqrt{\frac{2 E}{\mu}}
\end{eqnarray}
Because of the conservation of orbital angular momentum the impact
parameter is given by
\begin{eqnarray}
b=\frac{L}{\mu V_{-\infty}}=\frac{L}{\mu \sqrt{2 E/\mu}}.
\end{eqnarray}
Figure  \ref{ecc_6} shows the probability distribution function of 
impact parameters 
which shows a peak around $b \approx 0.5$. 
 We fit the distribution with a function of the form:
\begin{eqnarray}\label{fit}
  \frac{dP}{dx}dx&=&\frac{1}{a_2}\frac{a_0}{a_1} \left(\frac{x}{a_1} \right)^{a_0-1}
\exp \left(-\frac{x}{a_1}  \right)^{a_0} dx,
\end{eqnarray}
and we present the best fit parameters in table \ref{tab}. 
In the upper graph of fig. \ref{ecc_7} we show the correlation between the
impact parameter and 
the pericenter distance. The line is a power-law fit to the data with  
\begin{eqnarray}
r_{peri}=0.3 b^{0.72}. 
\end{eqnarray}
 The data shows a large scatter which makes the power-law fit a rather 
poor fit. However, to guide the eye and show the general 
trend  we include this fit.
The lower graph of fig. \ref{ecc_7} displays the
correlation between eccentricity and impact parameter of the encounter. Again
it becomes evident that the many of the orbits are close to be parabolic
and that only a small fraction is significant different from
parabolic. Encounters having $e>1$ merge very slowly if at all. This explains why
only hyperbolic encounters with small impact parameters $b < 1$ Mpc, 
corresponding to a close flyby,  lead to mergers.\\

\begin{figure}
\begin{center}
\includegraphics[width=8cm,angle=0]{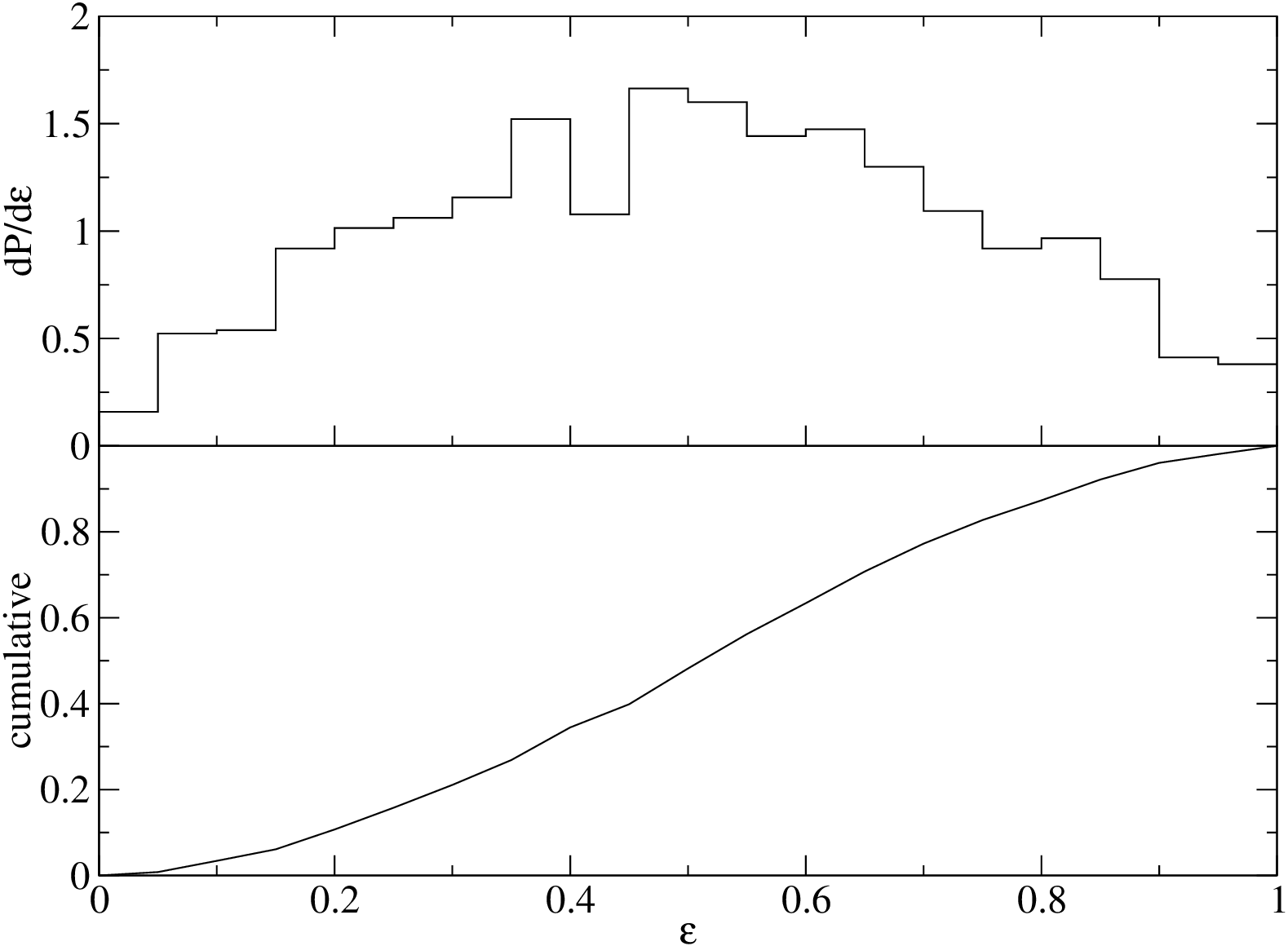}
\medskip
\caption{ Upper graph: Probability density of
 bound orbits with different circularity.
 Lower graph: To the upper graph corresponding cumulative fraction.
  Results are shown for progenitor halos of mass larger than  
 \mbox{$10^{12} M_{\odot}$} and mergers with mass ratio $M_h /M_s
 \leq 4$.  
  \label{ecc_9}}   
\vspace{1cm}
\includegraphics[width=8cm,angle=0]{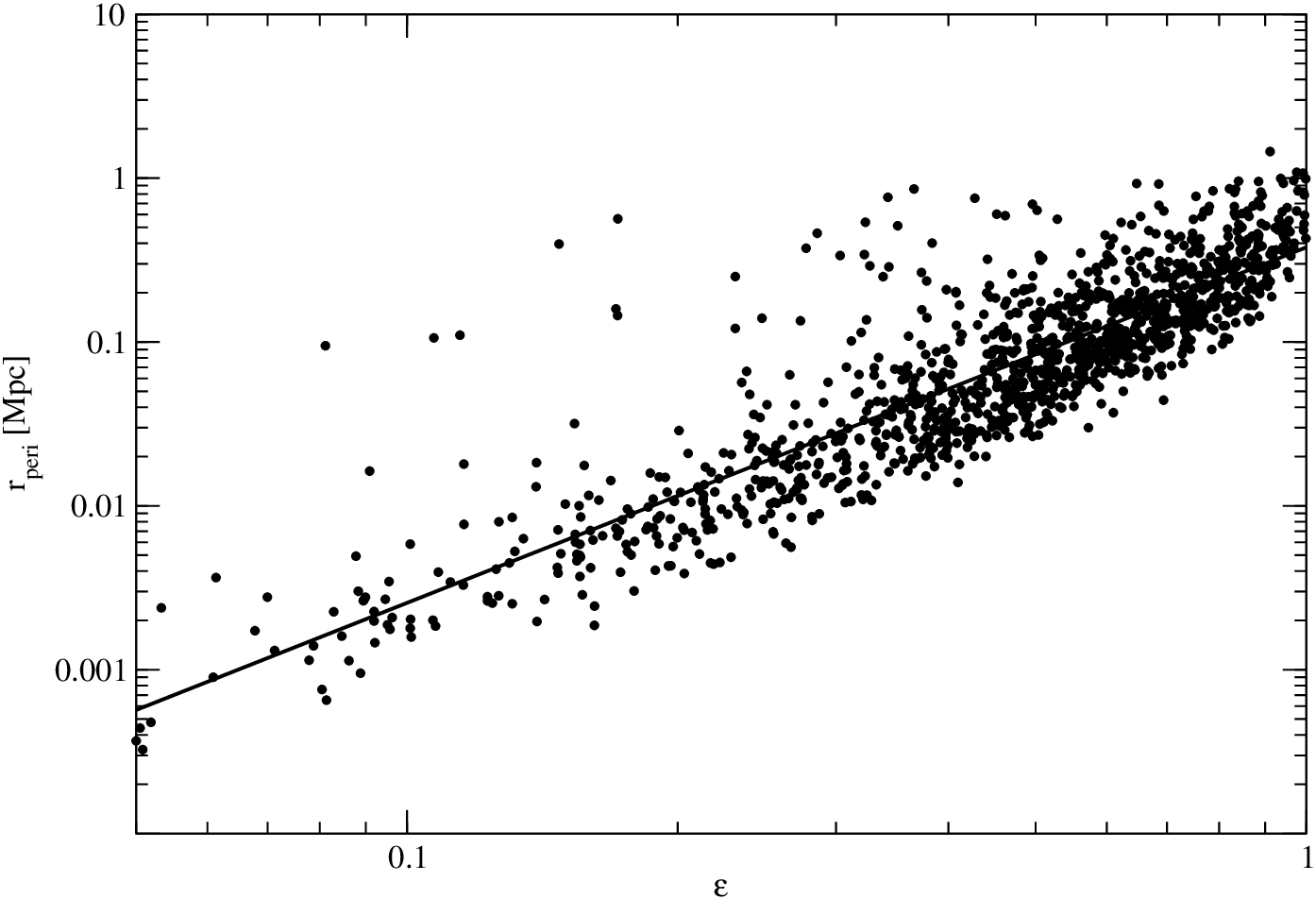}
\medskip
\caption{ Correlation between pericenter distance and circularity
 of bound orbits leading to mergers.
 Results are shown for progenitor halos of mass larger than  
 \mbox{$4 \times 10^{12} M_{\odot}$} and mergers with mass ration $M_h /M_s
 \leq 4$.  
  \label{ecc_8}}   
\end{center}
\end{figure}

Another parameter commonly used to describe bound orbits with $E < 0$ is the
circularity $\epsilon$ defined as the ratio of 
the orbital angular momentum to the angular momentum of a circular orbit with
the same energy. The circularity of a bound orbit can be derived applying the
virial theorem $U=-2T$. With
\begin{eqnarray}
r_{circ}=\frac{G M_h M_s}{2 E}, \quad V_{circ}=\sqrt{\frac{-2 E}{\mu}} \label{circu} 
\end{eqnarray}
and the angular momentum 
\begin{eqnarray}
L_{circ}=r_{circ} \mu V_{circ},
\end{eqnarray}
the circularity becomes
\begin{eqnarray}\label{circ_2}
\epsilon=\frac{L}{L_{circ}}=\frac{L}{r_{circ} \mu V_{circ}}.
\end{eqnarray}
From eq. \ref{circu} one sees that the circularity can only be  defined
for orbits with $E < 0$. Manipulating the appropriate equations gives the 
following relation for the circularity and eccentricity of an orbit:
\begin{eqnarray}\label{circ_3}
\epsilon=\sqrt{1-e^2}.
\end{eqnarray}
The upper graph of fig. \ref{ecc_9} presents the distribution of circularities
found. This result agrees with that of \citet{tor97}
who found the circularities to be distributed with a peak around $\epsilon \approx 0.5$.
However, in their analyses they considered only minor mergers $M_h/M_s \gg 4$ in a cluster environment,
where  the enhanced gravitational field might lead to a change of the circularity distribution. Our 
results show that these effects seem to play not an important role, 
and that the circularity distribution seems to be
 universal. An important consequence for semi-analytic modelling is that  
the circularity $\epsilon$ used to calculate the dynamical friction time scale
for major mergers must not be drawn from a
uniform distribution as e.g. done in \citet{kau99} 
 but from the distribution found here. 

Fig. \ref{ecc_8} shows that the correlation between 
circularity and the pericenter distance.
We find that the data can be fitted well by following power-law
\begin{eqnarray}
r_{peri}=0.38 \epsilon^{2.17} \mbox{Mpc}
\end{eqnarray}

\section{Parameters $\omega$ \& $i$}  
\begin{figure}
\begin{center}
\includegraphics[width=8.cm,angle=0]{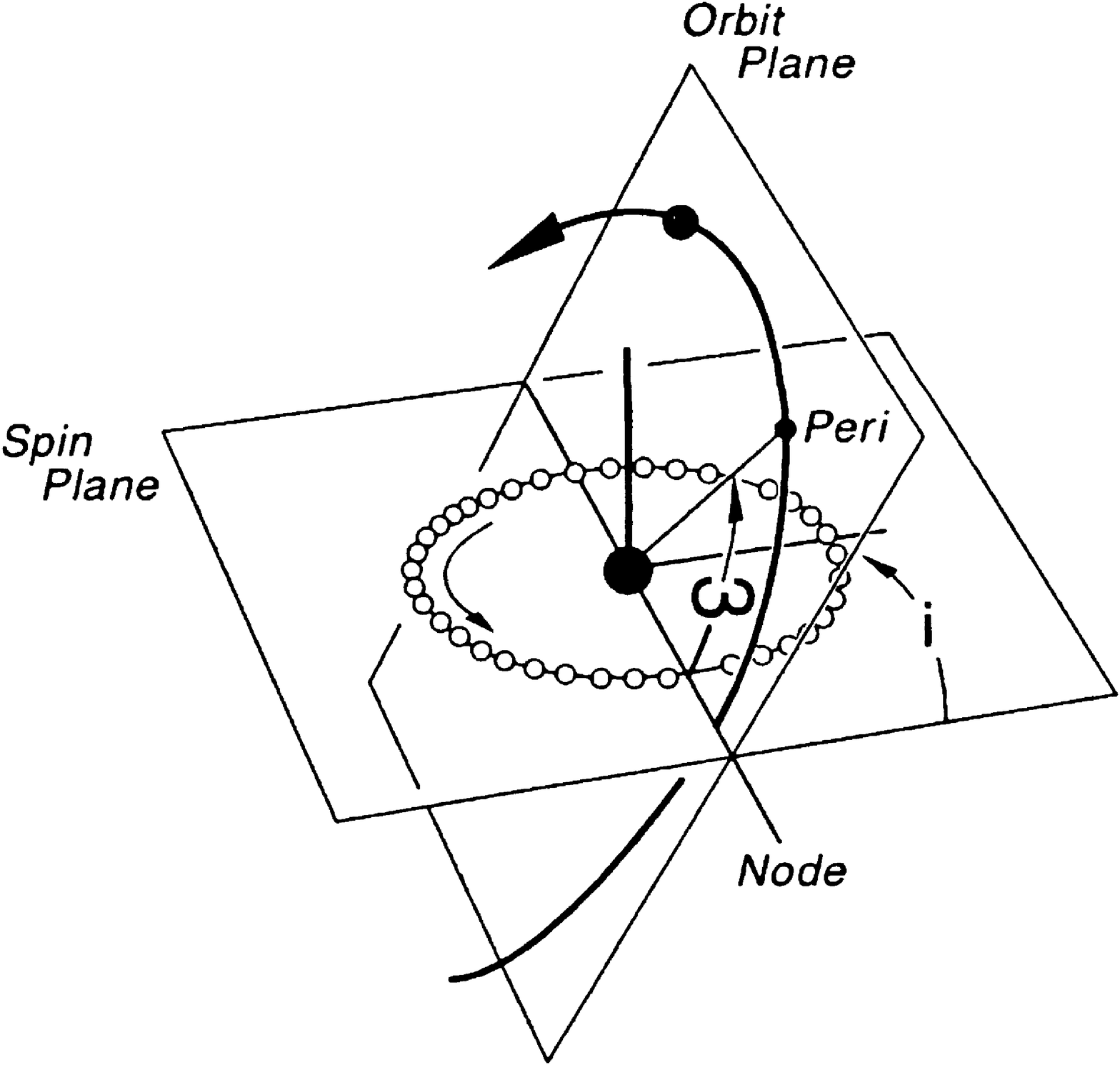}
\medskip
\caption{Definition  of the angles $i$ and $\omega$ following TT72
  (figure reproduced with permission of A. Toomre).
  \label{tt}}    
\end{center}
\end{figure}

If halos have spin, the orbital parameters discussed in the last section are
 not enough to fully
describe the geometry of the encounter. Additional constraints on the position
of the spin vectors ${\bf S}$ of both halos are required. These spin vectors are
 in general expected to align
with the spin vectors of the galactic disks that form by gas infall and kinetic 
energy dissipation. In calculating the spin vectors we follow \citet{lk99} who derived 
spin parameters from a similar GIF simulation. They found that for halos with more than 
70 particles halo properties are calculated reliably. Additionally they 
require that between 35 and 95 per cent of the friends-of-friends mass of the halo are inside 
of the virial radius. They find that 80 per cent of the halos satisfy these requirements.
In our analysis we adopt  limits of 50 particles and more
 per halo and between 50 and 95 per cent of the friends-of-friends mass inside the virial 
radius. In that way we make sure that we get results which are not influenced by ambiguities 
in the way we treat the particles inside each halo. In fig. \ref{tt} the definition of
the two corresponding angles is shown. The angle $i$ is defined in the rest frame
of the halo as the angle between the spin plane of the halo and the orbital
plane of the satellite; and in the rest frame of the satellite as the angle between the spin
plane of the satellite and the orbital plane of the halo. 
 By spin and orbital plane we refer to the planes perpendicular 
to the spin and orbital angular momentum vector, respectively.
These two angles $i_h$ and
$i_s$ are independent and by definition $|i| \le 180^{\circ}$, where
$i=0^{\circ}$ is a
prograde and $i=180^{\circ}$ a retrograde encounter. Additionally, the pericentric
argument $\omega$, is defined as the angle between the line of nodes and
separation vector at pericenter, and has values ranging from
$\omega=-90^{\circ}$ to
$\omega=90^ {\circ}$. It is not defined for $i=0^{\circ}$ or
$i=180^{\circ}$.  

\begin{figure}
\begin{center}
\includegraphics[width=8cm,angle=0]{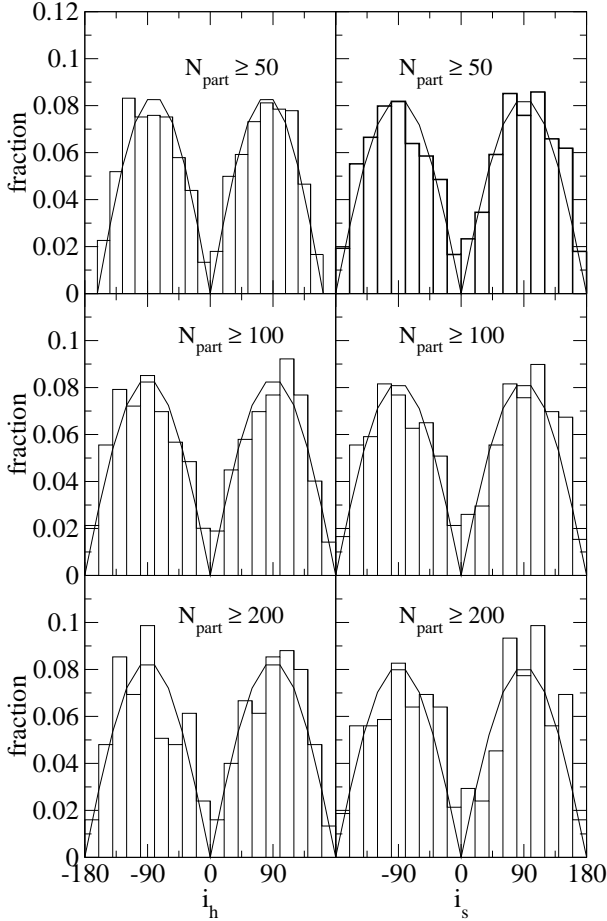}
\medskip
\caption{Left column: Angle between halo spin plan and orbital plane for
  different choices of minimum progenitor mass and a major merger definition of
  $M_h / M_s 
  \leq 4$. Right column: Same as left column but now for the angle between the
  spin plane of the satellite and orbital plane.  
  \label{ecc_10}}   
\end{center}
\end{figure}

\begin{figure}
\begin{center}
\includegraphics[width=8cm,angle=0]{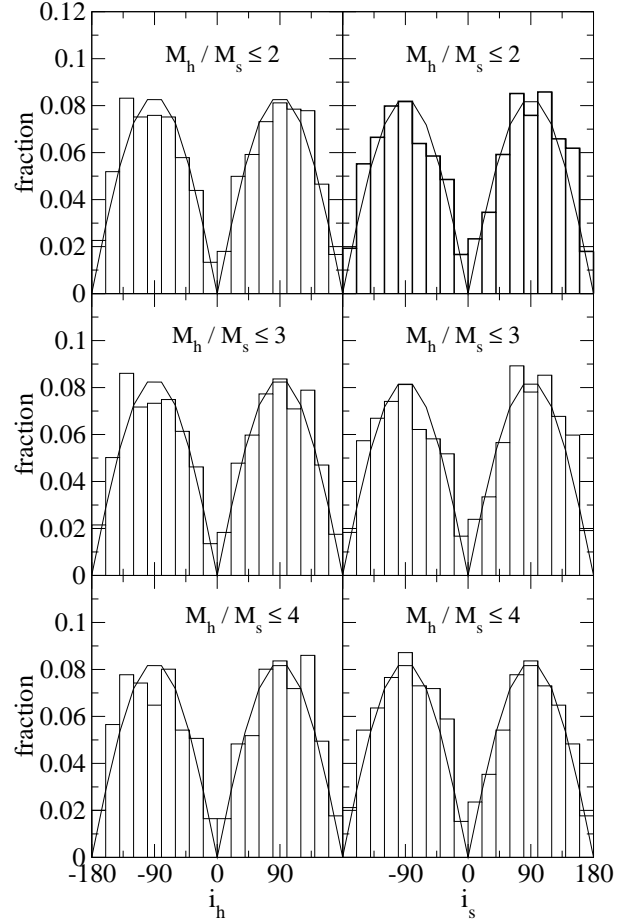}
\medskip
\caption{ Left column: Angle between halo spin plan and orbital plane for
  different choices of major merger definition and fixed minimum progenitor
  mass of $10^{12} M_{\odot}$. Right column: Same as left column but
  now for the angle between the 
  spin plane of the satellite and orbital plane.  
  \label{ecc_11}}   
\end{center}
\end{figure}

\begin{figure}
\begin{center}
\includegraphics[width=8cm,angle=0]{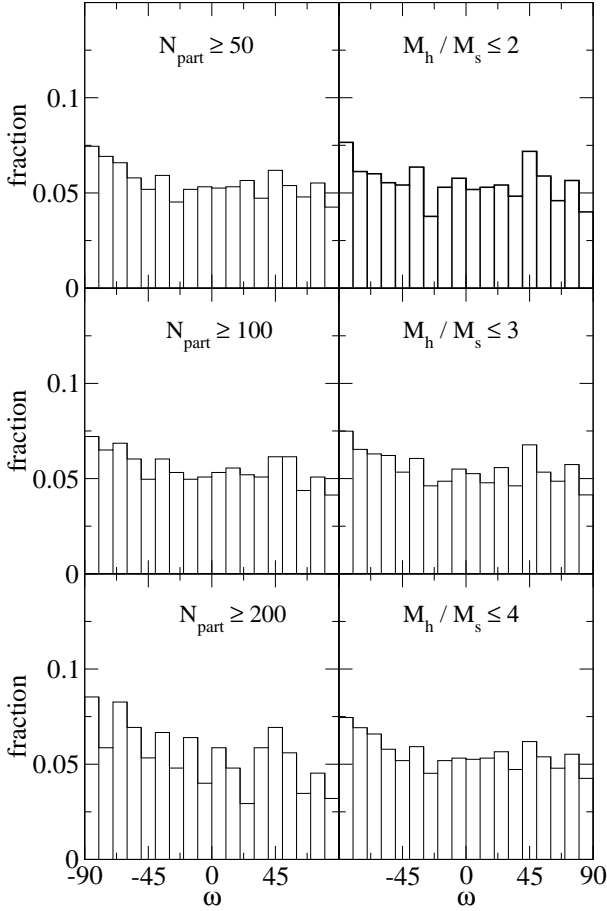}
\medskip
\caption{Left column: Angle between pericenter vector and node line  for
  different choices of minimum progenitor mass and a major merger definition of
  $M_h / M_s 
  \leq 4$. Right column: Same as left column but now for fixed minimum
  progenitor mass of  $ 10^{12} M_{\odot}$ and different
  choices of major merger definition. Results are the same for
  $\omega_s$ and $\omega_h$.
  \label{ecc_12}}   
\end{center}
\end{figure}

Figures \ref{ecc_10} - \ref{ecc_12} show the correlation between the angles $i$ and $\omega$ and
the minimum mass of progenitor halos or the mass ratios of the mergers.
The angles $i_h$ and $i_s$ are distributed following a sinus function.
The fit gets naturally
poorer at high minimum masses because of the smaller sample of halos.
The solid lines in figure \ref{ecc_10} and \ref{ecc_11} are
fits of the form $\propto |sin(x)|$. If the angle between two vectors is
sinus-distributed, the two vectors have no correlation. This can be understood
from looking at the probability of drawing a random vector pointing from the
centre of a sphere to its surface. If every point on the surface is equally
likely, the probability of finding an angle $i$ for example
between the x-axis and a random vector will be proportional to $\sin(i)$. We
therefore conclude that the spin plane and the orbital angular momentum plane
have no correlation.  Inspection of figure \ref{ecc_12} reveals that
the angle $\omega$ is equally distributed and independent of the
minimum halo mass and major merger definition. In contrast to $i$, $ \omega$
is defined by the position of a vector lying in a predefined plane. As a 
result an equal distribution in $\omega$ means that this angle is randomly
distributed. 

\begin{figure}
\begin{center}
\includegraphics[width=8cm,angle=0]{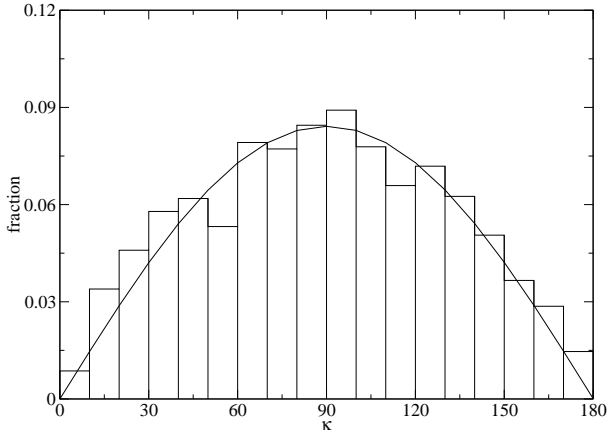}
\medskip
\caption{The distribution of angles $\kappa$ between the two spin planes.
  Results are shown for progenitor halos of mass larger than   
 \mbox{$ 10^{12} M_{\odot}$} and mergers with mass ratio $M_h /M_s
 \leq 4$.  
  \label{ecc_13}}   
\end{center}
\end{figure}

Defining $\kappa$ as the angle between the spin planes, one would expect from the
results presented above  that the spin vectors are also
not correlated with each other, and  that $\kappa$ should be sinus-like distributed.
The distribution of $\kappa$ is shown in  fig. \ref{ecc_13} and is indeed sinus
like. \\

\begin{figure}
\begin{center}
\includegraphics[width=8cm,angle=0]{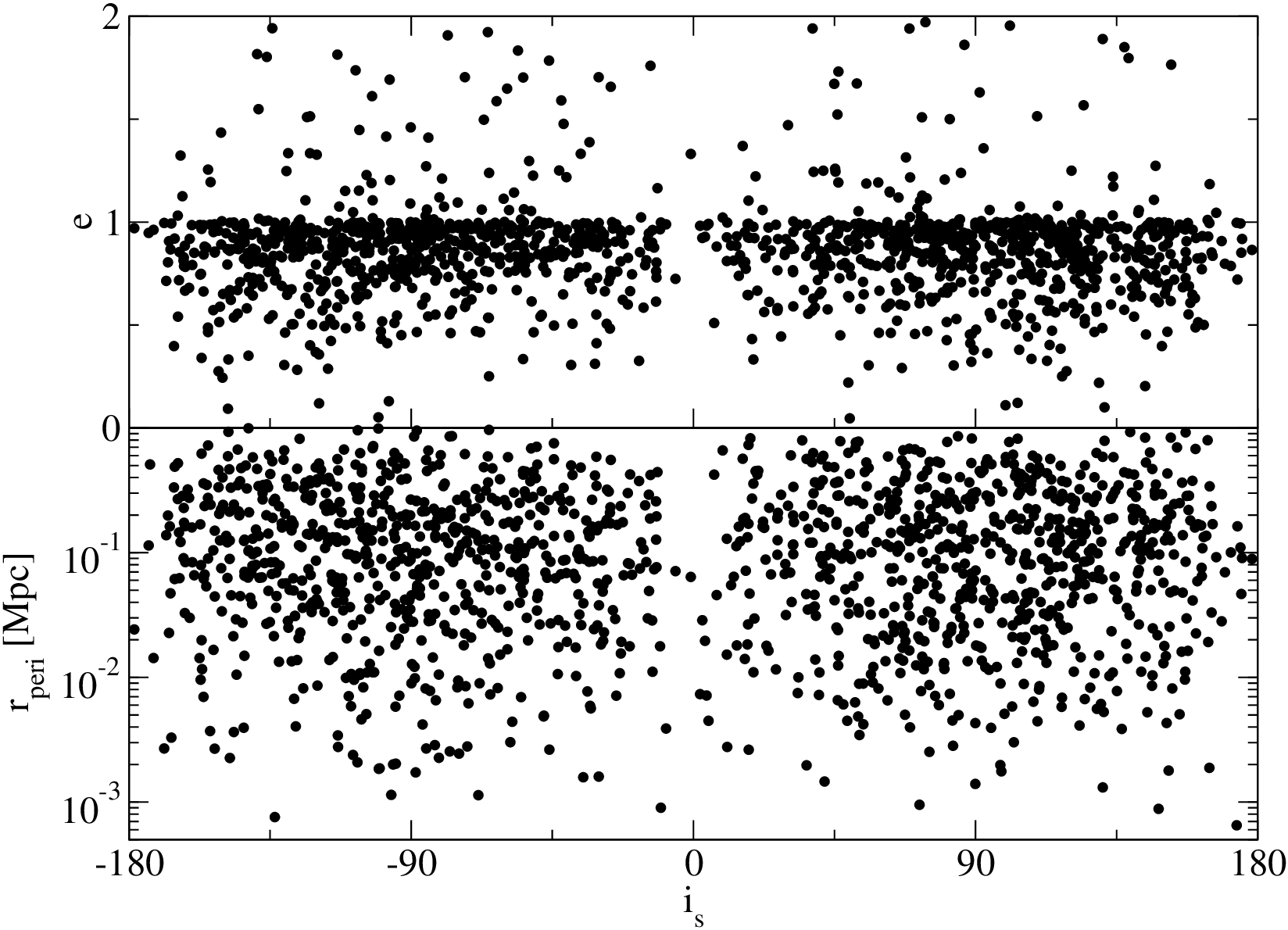}
\medskip
\caption{ Upper graph: Correlation between  $i_s$ of
  the satellite and the  eccentricities of 
 orbits leading to mergers. Lower graph: Correlation
 between the $r_{peri}$  and $i_s$ corresponding to the orbits in
 the upper graph.  Results are shown for progenitor halos of mass larger than  
 \mbox{$ 10^{12} M_{\odot}$} and mergers with mass ratio $M_h /M_s
 \leq 4$.  
  \label{ecc_14}}   
\end{center}
\begin{center}
\includegraphics[width=8cm,angle=0]{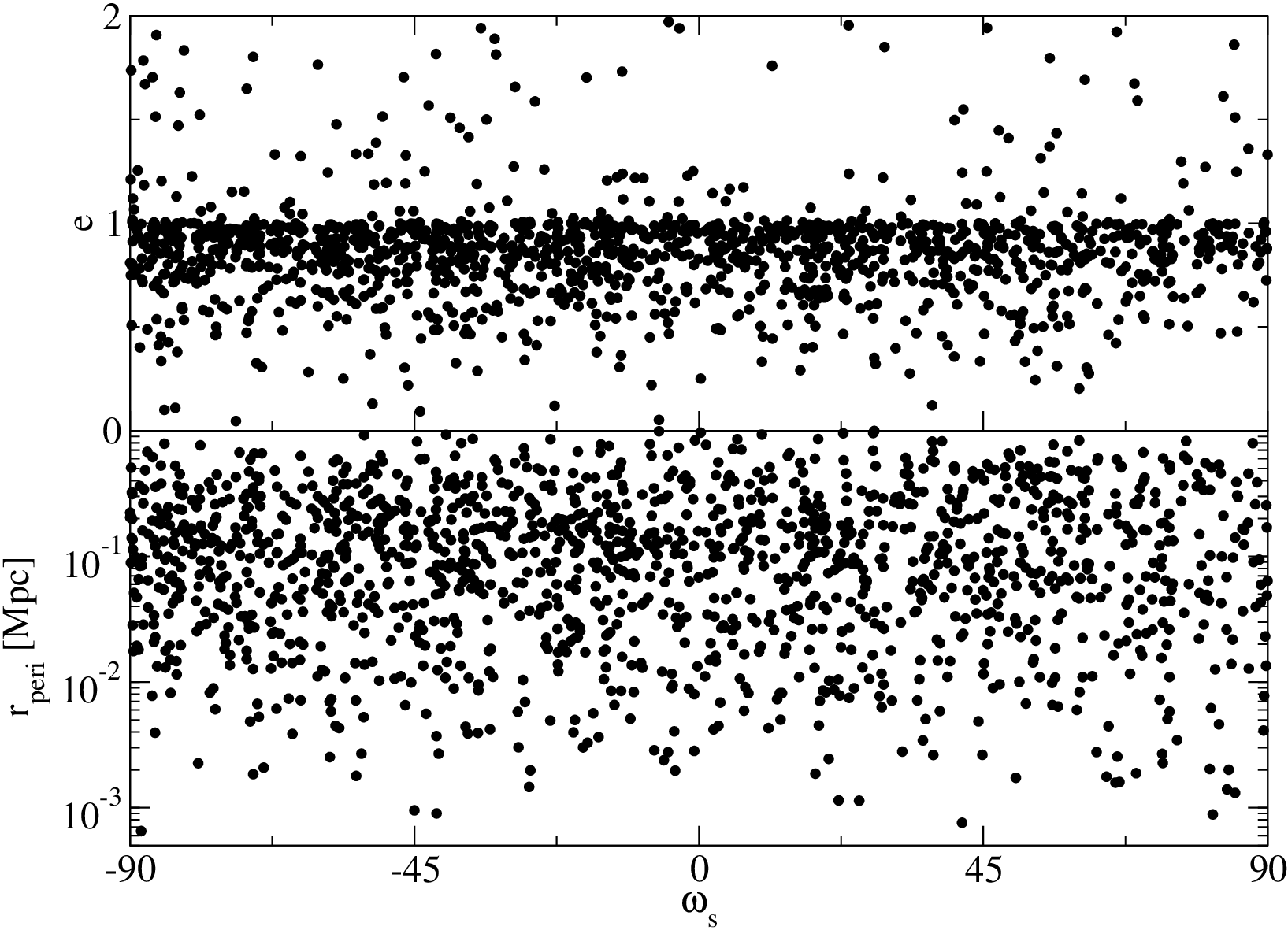}
\medskip
\caption{ Upper graph: Correlation between $\omega_s$ and $e$ for  orbits
  leading to mergers. Lower graph: Correlation 
 between $r_{peri}$ and $\omega_s$, corresponding to the orbits in
 the upper graph.  Results are shown for progenitor halos of mass larger than  
 \mbox{$ 10^{12} M_{\odot}$} and mergers with mass ratio $M_h /M_s
 \leq 4$.  
  \label{ecc_15}}   
\end{center}
\end{figure}

It is important to investigate possible  correlations between the
orbital parameters $r_{peri}$ and $e$ and the angles introduced
above. Fig. \ref{ecc_14} illustrates the correlation between the angle $i_s$
and the orbital parameters. The orbital parameters are not correlated 
 with $i_s$. 
 The same results are found for $i_h$. The pericentric 
argument $\omega_s$ shows no correlation with the orbital parameters as
shown in fig.  \ref{ecc_15}. The results are the same for $\omega_h$.   \\

\begin{figure}
\begin{center}
\includegraphics[width=8cm,angle=0]{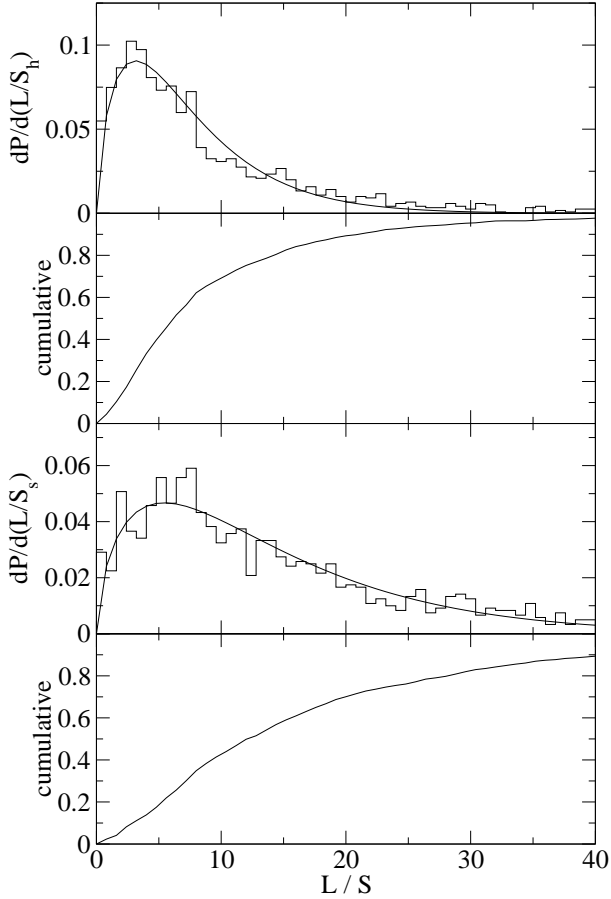}
\medskip
\caption{ The upper two graphs show the distribution
 and corresponding cumulative
  fraction of mergers with different $L/S_h$. The solid line in the upper of
  the two graphs is the fit using eq. \ref{fit}. Lower two graphs show the
  same as the upper graphs but now for the fraction $L/S_s$. 
   Results are shown for progenitor halos of mass larger than  
 \mbox{$ 10^{12} M_{\odot}$} and mergers with mass ratio $M_h /M_s
 \leq 4$.  
  \label{ecc_16}}   
\end{center}
\end{figure}

Semi-analytic models
\citep{mal02,vit02}  assume that during  mergers the orbital angular
momentum gets transformed into spin of the remnant halo. These models and 
models in which angular momentum is acquired by tidal torques e.g. Porciani, 
Dekel \& Hoffmann (2002)
reproduce the spin distribution of halos found in N-body simulations. However
the merger picture for the build up of halo spins uses some assumptions which
still need
 confirmation by N-body simulations. As a first step the amount of
angular momentum in the orbit must be investigated. Fig. \ref{ecc_16} shows
the distribution of the fraction of orbital angular momentum to spin of the
 halos $S_h$ and spin of the satellite $S_s$. 

\begin{figure}
\begin{center}
\includegraphics[width=8cm,angle=0]{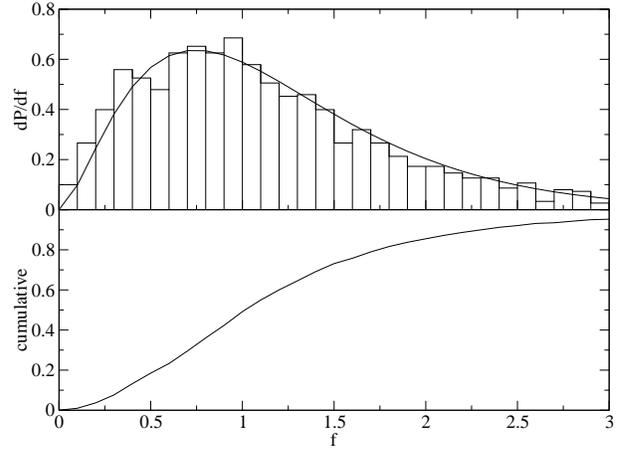}
\medskip
\caption{ Upper graph: Distribution of parameter $f$ found in the simulations.
 Lower graph: Corresponding cumulative fraction found in the simulation.
Results are shown for progenitor halos of mass larger than  
 \mbox{$10^{12} M_{\odot}$} and mergers with mass ration $M_h /M_s
 \leq 4$.  
  \label{ecc_17}}   
\end{center}
\end{figure}

\citet{mal02} define a parameter $f$ for mergers
\begin{eqnarray}
f=\frac{L}{V_{vir,h}r_{vir,h} \mu}
\end{eqnarray}
with $V_{vir,h}$ as the circular velocity 
 of the more massive progenitor. The value of this
parameter is set to be $ f \sim 0.42$ for their model, in which spin is
acquired from orbital angular momentum.
 In Fig. \ref{ecc_17} the distribution of $f$ is
displayed. 
 We fit the distributions in Fig. \ref{ecc_16} \& \ref{ecc_17} 
with functions of the form presented in eq. \ref{fit},
and the fitting parameters can be found in table \ref{tab}.
\begin{table}
\begin{tabular}{c|ccc}
\hline \noalign{\smallskip} \hline \noalign{\smallskip} 
$x$ & $a_0$ & $a_1$ & $a_2$  \\
\noalign{\smallskip} \hline \noalign{\smallskip} 
$b$ & 2.636 & 0.753 & 0.326 \\
$L/S_h$ & 1.727 & 7.408 & 0.662\\
$L/S_s$ & 1.624 & 14.132 & 0.726\\
$f$ & 2.638 & 1.2 & 0.307\\
\noalign{\smallskip} \hline
\end{tabular}
\caption{Best-fit parameters using eq. \ref{fit} to the distributions 
shown in Fig. \ref{ecc_6}, \ref{ecc_16} \& \ref{ecc_17}. \label{tab}}
\end{table}
The distribution for $f$ peaks at $\sim 0.6$ and has its 
mean at $ \sim 1.$ which
disagrees with the value required in the model of \citet{mal02}.  

\section{Discussion and Conclusions}  

We have investigated the orbital parameters of merging dark matter halos in a
cosmological large scale simulation  finding that the parameters used
generally in simulations of merging galaxies are in agreement with those found
in the large scale simulation. However, parameters used in merger
simulations up to now only occupy a very small part of the region of parameter space
covered by the results  from the large scale simulation. In particular
in most cases the calculated pericenter distances are more than a factor two
larger than those used generally in merger simulations. First results from
simulations using larger pericenter distances indicate more isotropic
remnants, which agree better with observations (Naab \& Burkert, in
preparation). About half of the orbits
are close to parabolic and those which are not show larger pericenter
distances on average. 
 The tight correlation $r_{peri} \propto \epsilon^{2.17}$ which we find
 between the pericenter distance and the circularity just
 supports that we have chosen halos when they are still well separated and not
 much interaction has taken place. Interaction leads to a circularization of
 the orbit predicting an increase of the circularity $\epsilon$ on average 
with decreasing pericenter distance $r_{peri}$. On the other hand hyperbolic
orbits need to have very high kinetic energies to allow for small pericenter
distances which the simulation shows to be unlikely. The pericenter distances
however, show in the case of hyperbolic orbits a correlation with the
impact parameter of the form $r_{peri} \propto b^{0.72}$, allowing to 
disentangle the different contributions to the orbital angular momentum using
the additional correlation $L_{sp} \propto r_{peri}^{0.53}$.

The distribution of circularities found is in contrast to what is commonly
assumed in semi-analytic models of galaxy formation, which assume a random
distribution of circularities for the orbits of infalling satellites. Instead
our results indicate a distribution with a peak around $\epsilon \approx 0.5$.
   
We investigated the correlation between the spin plane of the main halo and the
spin plane of the satellite with the orbital plane of the merger, finding no
correlation. The planes  are aligned randomly which gives
statistical weight to certain merger geometries. Additionally the spin planes
are not correlated with each other and the angle $i$ between the spin planes 
and the orbital plane shows no correlation with the other orbital parameters
$r_{peri}$ and $e$ which justifies its use as an independent initial parameter.
The  pericentric argument $\omega$ follows a random distribution and
shows no correlation with the other orbital parameters
$r_{peri}$ and $e$.

We tested the requirement needed in the semi-analytic model of \citep{mal02}
for the acquiring of angular momentum of halos 
through major mergers. The
authors define a parameter $f$ which should have on average a value of $f \sim
0.42$ to make their model reproduce the distributions found in
simulation. However we find a mean value $f \sim 1.$ which is larger than
that required by their model, suggesting an inconsistency.    

All presented results are equally valid for major mergers ($M_{1}/M_{2} \leq
4, M_{1} \geq M_{2}$) and show no dependence on the mass ratio or
mass of the merging
objects supporting a self similar build up of structure. The use of the
self-consistent initial conditions presented here will allow to test
consistently the merger scenario for the formation of elliptical galaxies in the
hierarchical universe without covering the whole parameter space and provides
appropriate statistical weights for different initial conditions. 
\newline
\begin{acknowledgements}
The authors would like to thank Thorsten Naab and Simon White for useful
comments and Hugues Mathis for his help dealing with the simulation outputs. We would also like to thank 
Andrew Benson for intensive discussion and pointing out the importance of the Hubble flow for the eccentricities of 
orbits. SK acknowledges support by PPARC Theoretical Cosmology Rolling Grant. 
\end{acknowledgements}



\end{document}